\documentclass[12pt]{article}

\usepackage{algpseudocode}
\usepackage{amsmath, amssymb, amsthm}
\usepackage{float, fullpage, graphicx, multirow,parskip, subcaption, setspace}
\usepackage{comment}
\usepackage{url}
\usepackage{enumitem}
\usepackage{tikz}
\usepackage{bm, bbm}
\usetikzlibrary{shapes.geometric, positioning}
\usetikzlibrary{quotes, angles}
\usepackage{rotating}
\usepackage{hyperref}
\usepackage{natbib}
\usepackage{placeins}
\usepackage{algorithm2e}
\RestyleAlgo{ruled}
\usepackage[english]{babel}
\usepackage{natbib}

\usepackage{lineno}
\usepackage{amsmath}
\usepackage{graphicx}

\usepackage{authblk}
\usepackage{bm}

\DeclareMathOperator{\Bernoulli}{Bernoulli}
\DeclareMathOperator{\logit}{logit}
\DeclareMathOperator{\Poisson}{Poisson}

\newcommand{\activitycenterplain}{s}
\newcommand{\activitycentervec}[2]{\bm{\activitycenterplain}_{#1,#2}}
\newcommand{\activitycenterveccomp}[3]{\activitycenterplain_{#1,#2}^{[#3]}}

\newcommand{\mvnormal}[2]{\mathcal{N}\left(#1, #2\right)}

\newcommand{\detectionrate}[3]{r_{#1,#2,#3}} 

\usepackage{setspace} 

\title{Explicit modeling of density dependence in spatial capture-recapture models}
\author[1]{Qing Zhao}
\author[2]{Yunyi Shen}
\affil[1]{Bird Conservancy of the Rockies, whitelangur@gmail.com}
\affil[2]{Department of Electrical Engineering and Computer Science, Massachusetts Institute of Technology, yshen99@mit.edu}

\usepackage[normalem]{ulem}

\newcommand{\delete}[1]{}

\newcommand{\revision}[1]{{ #1}} 
 
\newcommand{\revisions}[2]{{#1}}

\begin{document}
\maketitle

\begin{abstract}
Density dependence occurs at the individual level and thus is greatly influenced by spatial local heterogeneity in habitat conditions. However, density dependence is often evaluated at the population level, leading to difficulties or even controversies in detecting such a process. Bayesian individual-based models such as spatial capture-recapture (SCR) models provide opportunities to study density dependence at the individual level, but such an approach remains to be developed and evaluated. In this study, we developed a SCR model that links habitat use to apparent survival and recruitment through density dependent processes at the individual level. Using simulations, we found that the model can properly inform habitat use, but tends to underestimate the effect of density dependence on apparent survival and recruitment. The reason for such underestimations is likely due to the difficulties of the current model in identifying the locations of unobserved individuals without using environmental covariates to inform these locations. How to accurately estimate the locations of unobserved individuals, and thus density dependence, remains a challenging topic in spatial statistics and statistical ecology.

\end{abstract}

\section{Introduction}
Identifying the effect of density dependence in population regulation is a fundamental question in ecology and often relies on statistical approaches \citep{st1970detection,hanski1990density,turchin1992complex,hassell1975density,herrando2012density}. The answer to this question has broad implications for evolution and theoretical ecology \citep{bellows1981descriptive,ratikainen2008density,travis2013evolution,holt1996adaptive} and application values in predicting population viability and extension under habitat changes \citep{howell2020informing,zhao2019integrated,zabel2006interplay,sabo2004efficacy}. While most studies examined density dependence at the population level \citep{dennis1994density,brook2006strength,dennis2006estimating,hanski1990density}, processes that lead to density dependence, such as competition \citep{lamb2017influence,hixon2005competition,fronhofer2023shape}, often occur at the individual level and thus is greatly influenced by spatial local heterogeneity in habitat conditions. Such discrepancy can lead to difficulties in detecting density dependence and even controversial results and imposes challenges in statistical modeling \citep{detto2019bias,shenk1998sampling}.

Individual-based modeling has been used to understand how population-level patterns emerge from individual-level processes \citep{judson1994rise,grimm2013individual}. While individual-based modeling has traditionally relies on supervised learning \citep{grimm2013individual}, more recently researchers have emerged it with reinforcement learning under a Bayesian framework \citep[i.e., Bayesian individual-based modeling][]{hooten2020statistical,schafer2022bayesian}. Such a development provides opportunities to gain knowledge about individual-level processes, including density dependence across the individual and population levels in heterogeneous landscapes, with statistical modeling.

Spatial capture-recapture (SCR) was originally developed to estimate population size while assuming a ``selection-free'' kernel for habitat use \citep{efford2004density,borchers2008spatially,royle2008hierarchical,royle2013spatial,efford2025secr}. These models were then further extended to allow habitat use to be related to landscape characteristics at local scales \citep{royle2013integrating,royle2018unifying,zhao2024bayesian}. Dynamic SCR models allow vital rates (apparent survival, recruitment) to be related to population density, but often at the population scale \citep[but see \citealp{milleret2023estimating}]{gardner2010spatially,chandler2014spatially,ergon2014separating}. Because SCR models utilize individual-based data and explicitly consider the unique location, and thus environment, of each individual, there is a potential to relate vital rates to density dependence at the individual level in these models, along the lines of Bayesian individual-based modeling. However, SCR models that examine the effect of density dependence on vital rates at the individual level remains to be developed and evaluated.

Here, we developed a novel SCR model that estimates habitat use information and links such information to density dependent apparent survival and recruitment rates at the individual level. We evaluated the model with simulations and a case study, and discussed the pros and cons of our modelling approach and SCR models in general.

\section{Methods}

\subsection{Model description}
\begin{figure}[htp]
\centering
\includegraphics[width=1.0\linewidth]{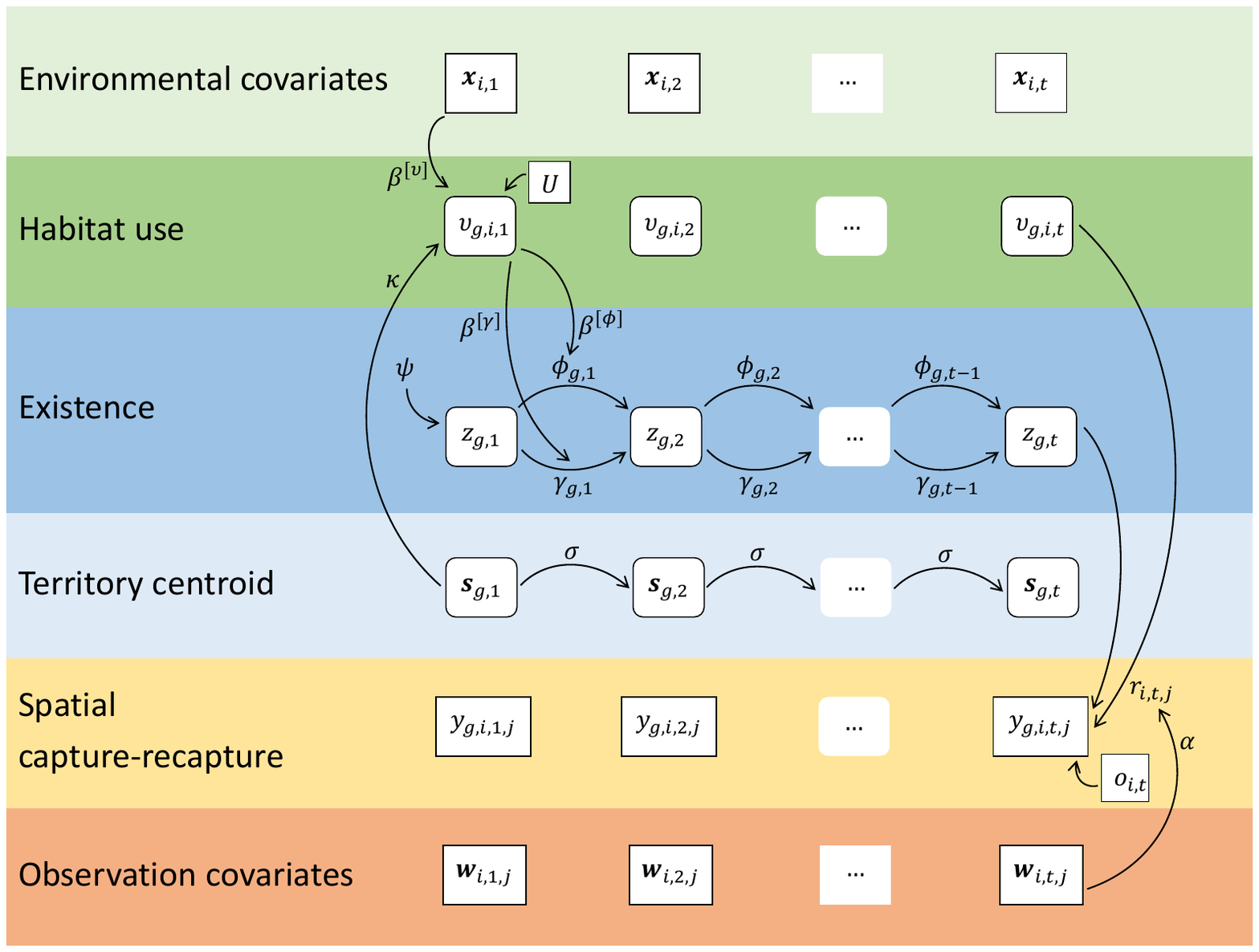}
\revision{\caption{\label{fig:diagram}Diagram showing the structure of the SCR model introduced in the current study. The model utilizes SCR data $y$, alongside with environmental covariate data $x$, observation covariate data $w$, and survey information $o$, given survey period $U$, to inform habitat use $\upsilon$ with the effect of environment $\beta^{[\upsilon]}$ and the effect of distance $\kappa$ on habitat use, individual existence status $z$ driven by the probabilities of initial existence $\psi$, apparent survival $\phi$ and recruitment $\gamma$ with the effects of habitat use on density dependent apparent survival $\beta^{[\phi]}$ and recruitment $\beta^{[\gamma]}$, the locations of territory centroids $s$ with the standard deviation of Gaussian dispersal $\sigma$, and detection rate $r$ with the effect of observation covariates on detection $\alpha$.} }
\end{figure}

We developed a dynamic spatial capture-recapture (SCR) model that contains four sub-models: a sub-model that describes habitat use, a sub-model that links habitat use with individual existence through apparent survival and recruitment, a sub-model that describes dispersal movement, and eventually an observation sub-model that link habitat use with frequency of detection data (Figure \ref{fig:diagram}).

The first sub-model describes the amount of time of habitat use by individual animals such that 
\begin{equation}
\begin{aligned}
    \upsilon_{g,i,t}= U \times \frac{\exp(-\kappa d_{g,i,t}+\beta^{[\upsilon]}_1 x_{i,t,1}+\ldots+\beta^{[\upsilon]}_{n_{covs}} x_{i,t,n_{covs}})}{\sum_{i=1}^{n_{site}}\exp(- \kappa  d_{g,i,t}+\beta^{[\upsilon]}_1 x_{i,t,1}+\ldots+\beta^{[\upsilon]}_{n_{covs}} x_{i,t,n_{covs}})}.
\end{aligned}
\end{equation}
Like other SCR models \citep[e.g.,][]{efford2004density, royle2018unifying}, we assumed that every individual has a territory with a centroid. However, our model is different from previous SCR models in modelling a continuous variable $\upsilon_{g,i,t}$, which represents the true but latent amount of time of individual $g$ using site $i$ \revision{during the primary period (e.g., breeding season)} in year $t$, and is assumed to be a function of the distance between individual $g$’s territory centroid and site $i$, denoted $d_{g,i,t}$, with a decay parameter $\kappa$, as well as $n_{covs}$ potentially time varying environmental covariates $x$’s, with corresponding parameters $\beta$’s representing the direction and strength of habitat selection. We also assumed that $\upsilon_{g,i,t}$ add up to $U$, a \textit{known} quantity for the total amount of time \revision{of the primary period} covered by the survey, under the assumption of closed population. To satisfy this assumption, the study area needs to be large enough and also relatively isolated so that it contains a distinguishable population, and there should be no changes in the population size due to birth, death, or dispersal movement within each primary period. To satisfy this assumption, researchers also need to model all sites that individuals in this population could possibly use, regardless if the sites are surveyed or not. To define these sites, researchers can either rely on prior knowledge about what a site is, or consider every spatial units (e.g., grid cells) in the study area as possible sites and let the model determine if they are used by the animals or not.

The second sub-model describes how the true amount of time of habitat use influences individual existence through density dependent apparent survival and recruitment such that the status of an individual $g$ in year $t$ $z_{g,t}$ (0 for not alive and 1 for alive) follows
\begin{align}
    & z_{g,t=1}\sim \Bernoulli(\psi) \\
    & z_{g,t\geq2}\sim \Bernoulli\left[z_{g,t-1}  \phi_{g,t-1}+(1-e_{g,t-1}) \gamma_{g,t-1}\right] \\
    & \phi_{g,t}=\logit^{-1}\left[\beta^{[\phi]}_0+\beta^{[\phi]}_{density} \times \frac{\sum_{i=1}^{n_{site}}\upsilon_{g,i,t}\left(\sum_{k:k\ne g}\upsilon_{k,i,t} z_{k,t}\right) }{\sum_{i=1}^{n_{site}}\upsilon_{g,i,t}}\right] \\
    & \gamma_{g,t}=\logit^{-1}\left[\beta^{[\gamma]}_0+\beta^{[\gamma]}_{density} \times \frac{\sum_{i=1}^{n_{site}}\upsilon_{g,i,t}\left(\sum_{k:k\ne g}\upsilon_{k,i,t} z_{k,t}\right) }{\sum_{i=1}^{n_{site}}\upsilon_{g,i,t}}\right]
\end{align}
in which $z_{g,t}$ represents the status of individual $g$ in year $t$ such that $z_{g,t}=1$ if individual $g$ exists (i.e., is alive) in year $t$ and $z_{g,t}=0$ otherwise; $\psi$ is the probability of existence in the first year; from the second year, each individual has a probability of $\phi_{g,t-1}$ to survive given it is alive in the previous year (i.e., $z_{g,t-1}=1$), and a probability of $\gamma_{g,t-1}$ to be recruited to the population given it hasn’t existed so far (i.e., $e_{g,t-1}=0$); $e_{g,t}$ is an indicator which equals 1 if individual $g$ has been alive in any of the years up until year $t$, and 0 otherwise; $\phi_{g,t-1}$ and $\gamma_{g,t-1}$ are functions of local density with intercept and slope parameters $\beta^{[\phi]}$’s and $\beta^{[\gamma]}$’s, respectively. \revision{Thus, while we assumed a closed population within each primary period, we allowed changes in the population size between primary periods, which is consistent with many other open-population models \citep{broms2016dynamic, efford2020spatial, howell2020informing, zhao2020sampling, milleret2023estimating}.} The key component of this model, which distinguishes it from other SCR models, is to aggregate habitat use by any individuals other than $g$, $\upsilon_{-g,i,t}$, given they are alive, i.e., $z_{-g,t}=1$, and calculate their weighted mean, using habitat use of individual $g$, $\upsilon_{g,i,t}$, as the weight, across all sites. This approach allows us to represent density dependent processes due to competition such that an individual that more frequently uses habitat sites that are less frequently used by other individuals tends to face a lower level of competition and thus has higher apparent survival and recruitment.

The third sub-model describes dispersal movement as an autoregressive process such that 
\begin{equation}
\activitycentervec{g}{t}\sim \mvnormal{\activitycentervec{g}{t-1}}{\sigma \bm I}
\end{equation}
in which $\activitycentervec{g}{t}=(\activitycenterveccomp{g}{t}{x},\activitycenterveccomp{g}{t}{y})$ is the two dimensional vector of individual $g$'s territory centroid in year $t$ with easting $\activitycenterveccomp{g}{t}{x}$ and northing $\activitycenterveccomp{g}{t}{y}$. \revision{Again, we assumed that the territory centroids are constant within each primary period, and only allowed the changes of their locations between primary periods.} Also, by having a two dimensional identity matrix $\bm I$ with the standard deviation of Gaussian dispersal movement $\sigma$ to represent an isotropic Gaussian dispersal kernel, we assumed that there is no preferred direction of dispersal. This assumption can be easily extended by allowing correlations between $\activitycenterveccomp{g}{t}{x}$ and $\activitycenterveccomp{g}{t}{y}$. 

Eventually, the last sub-model describes how detection data are obtained given the true amount of time of habitat use such that 
\begin{equation}
\begin{aligned}
    & y_{g,i,t,j}\sim \Poisson(\detectionrate{i}{t}{j}o_{i,t}z_{g,t}\upsilon_{g,i,t}    ) \\
    & \detectionrate{i}{t}{j} = \exp\left(\alpha_0+\sum_{k=1}^{n_{\mathrm{pcvs}}}\alpha_k w_{i,t,j,k}\right)
\end{aligned}
\end{equation}
in which discrete data $y_{g,i,t,j}$ represents the \textit{number} of times for which individual $g$ is captured at site $i$ in year $t$ by the $j$th camera, $\detectionrate{i}{t}{j}$ is the detection rate and is a function of $n_{\mathrm{pcvs}}$ observation covariates $w$’s with intercept and slope parameters $\alpha$’s, and $o_{i,t}$ indicates the proportion of time for which a camera trap is deployed at site $i$ in year $t$. Note that we assumed to have multiple camera traps deployed at each site to allow replicates of the observations \citep{pollock1982capture}; these camera traps need to function independently, meaning that capturing the animal by one camera trap should not influence the function of other camera traps. Also note that $\detectionrate{i}{t}{j}$ represents average number of detections, rather than a probability, when an activate individual spent unit time at a site that had unit survey effort $o_{i,t}$, and thus is not necessarily upper bounded by 1. Also, because we modelled all possible sites, $o_{i,t}$ allows the situation that camera traps are not necessarily deployed at every site and they can be deployed for different amount of time.

\subsection{Simulation design}
During the simulations, we simulated a total of 200 individuals, including individuals that actually existed and individuals that did not exist, meaning that we used data augmentation \citep{royle2007analysis,lewis2023climate}. \revision{More specifically, the parameter $\psi$ governs the number of individuals that actually exist in the first year of the study. In later years, some non-existing individuals can be recruited into the population, the probability of which is governed by $\gamma$. Eventually, there will be some individuals that never exist during the entire study period, and these individuals are "augmented". Being able to estimate the above-mentioned parameters indicates that data augmentation does not induce systematic bias when a proper number is selected for the augmented population size, shown by previous studies \citep{royle2007analysis, milleret2023estimating} and the current study (see results). In practice, we recommend selecting a number that is a few times of the potential population size. Selecting a small number, which will be indicated by an estimated $\psi$ that is close to 1, can lead to underestimation of population size. On the other hand, selecting an extremely large number can make the computation inefficient and lead to small $\psi$ and $\gamma$ values that are difficult to estimate.}

We simulated a landscape with 500 sites within a 20-unit wide and 16-unit long area during a 6-year period of time; the landscape was characterized by two spatially and temporally varying covariates, both of which were randomly generated from a standard Normal distribution. We assumed that territory centroids were uniformly distributed in the landscape in the first year. We set $\kappa$ to 2.5 to allow a moderate size of territory. We set $\beta^{[\upsilon]}_1$ to 0.6 and $\beta^{[\upsilon]}_2$ to -0.3 to allow opposite effects of the two covariates on habitat use. We set $\psi$ to 0.4, meaning that more than half of the individuals did not exist in the first year of the study. We set $\beta^{[\phi]}_0$ to 1 and $\beta^{[\gamma]}_0$ to -1 to allow a relatively high apparent survival and relatively low recruitment. We set $\beta^{[\phi]}_{density}$ to -0.4 and $\beta^{[\gamma]}_{density}$ to -0.6 to allow negative density dependence on both apparent survival and recruitment. We set $\sigma$ to 2.5 to allow a relatively short distance of dispersal movement. For the observation processes, we considered two observation covariates which again were randomly generated from a standard Normal distribution. We set $\alpha_0$, $\alpha_1$ and $\alpha_2$ to 0.7, -0.6, 0.3, respectively. For cameras in the middle part (16-unit wide and 12-unit long) of the study area, $o_{i,t}$ was randomly generated from a uniform distribution between 0.8 and 1; $o_{i,t}$ was set to 0 to represent the situation that cameras were not deployed outside the middle part of the study area.

We considered two models in our simulations. We first considered a simpler model which only allows density dependence on recruitment but not apparent survival. This is equivalent to have $\phi_{g,t}=\logit^{-1}(\beta^{[\phi]}_0)$ in equation 2. This model also does not allow dispersal movement between years, which mean we have $\activitycentervec{g}{t}=\activitycentervec{g}{}$ for any $t$. We then considered a more complex model that allows density dependence on both recruitment and apparent survival, and dispersal movement between years, as described above.

\subsection{Model evaluation}
We wrote our own MCMC algorithm in R \citep{r2024r}. We repeated the simulations for 20 times with newly generated data each time. We ran 50,000 iterations, including 25,000 burn-in, in 3 chains for each simulation. We pooled all posterior samples from all simulations to make violin plots, which also provided key statistics including median, 50\% Credible Interval (CI) and 95\% CI. We considered a parameter estimate to be biased if the 50\% CI does not cover the true value of the parameter used in the simulation.

\subsection{Real data analysis}
We applied our proposed method on \revisions{nine}{ten} years tiger survey data from India by \citet{karanth2006assessing}. The published version of the dataset by \citet{tigerdata} contained ten \revisions{primary periods}{years} of surveys, however the first two surveys were \revisions{conducted within the same year}{spaced only half a year}. We therefore \revisions{removed the data from the second survey to constrain the midpoints of all surveys in the first half of the years for consistency}{considered this data to have nine primary period excluding the first snapshot}. \revision{We delineated the study area into 342 grid cells that are 1 $km^2$ in size. When there were multiple (maximally 7) cameras deployed at the same sites, we considered them as spatial replicates. There were in total 74 individuals captured during these nine surveys, and we augmented our data to have 200 potential individuals. }\revisions{Survey effort, measured by days in each primary period a camera was active, ranged from 1 to 46 days. We divided these active days by the maximum active day (i.e., 46) to calculate the survey effort $o_{i,t}$, while using 46 as the time period covered by the survey $U$.}{Survey effort, measured by days in each primary period a camera was active, ranged from 33 to 197 days. We used these active days to calculate our effort data $o_{i,t,j}$ dividing days active by 365. We set total number of time covered by survey $U$ to be 1, as measured in year.}\revisions{}{There was no replicates within each primary period. There were in total 76 individuals seen thus we augmented our data to have 200 potential individuals.} The published version of the dataset does not contain any environmental variables\revisions{}{and the spatial locations are altered to conserve tiger locations which made it hard to obtain environmental variables through georeferencing}, therefore we did not include \revisions{any environmental covariates}{them} in our analysis. \revisions{}{Geometric locations were converted to have unit of kilometers instead of original meters. }We ran three chains with 50,000 iterations \revision{including 10,000 burn-in} in each chain. \revision{The convergence of the MCMC computing was checked using the diagnostic plot of the posterior samples and Brooks–Gelman–Rubin Diagnostics.}

\section{Results}
\subsection{Simulation}
\begin{figure}[htp]
\centering
\includegraphics[width=1.0\linewidth]{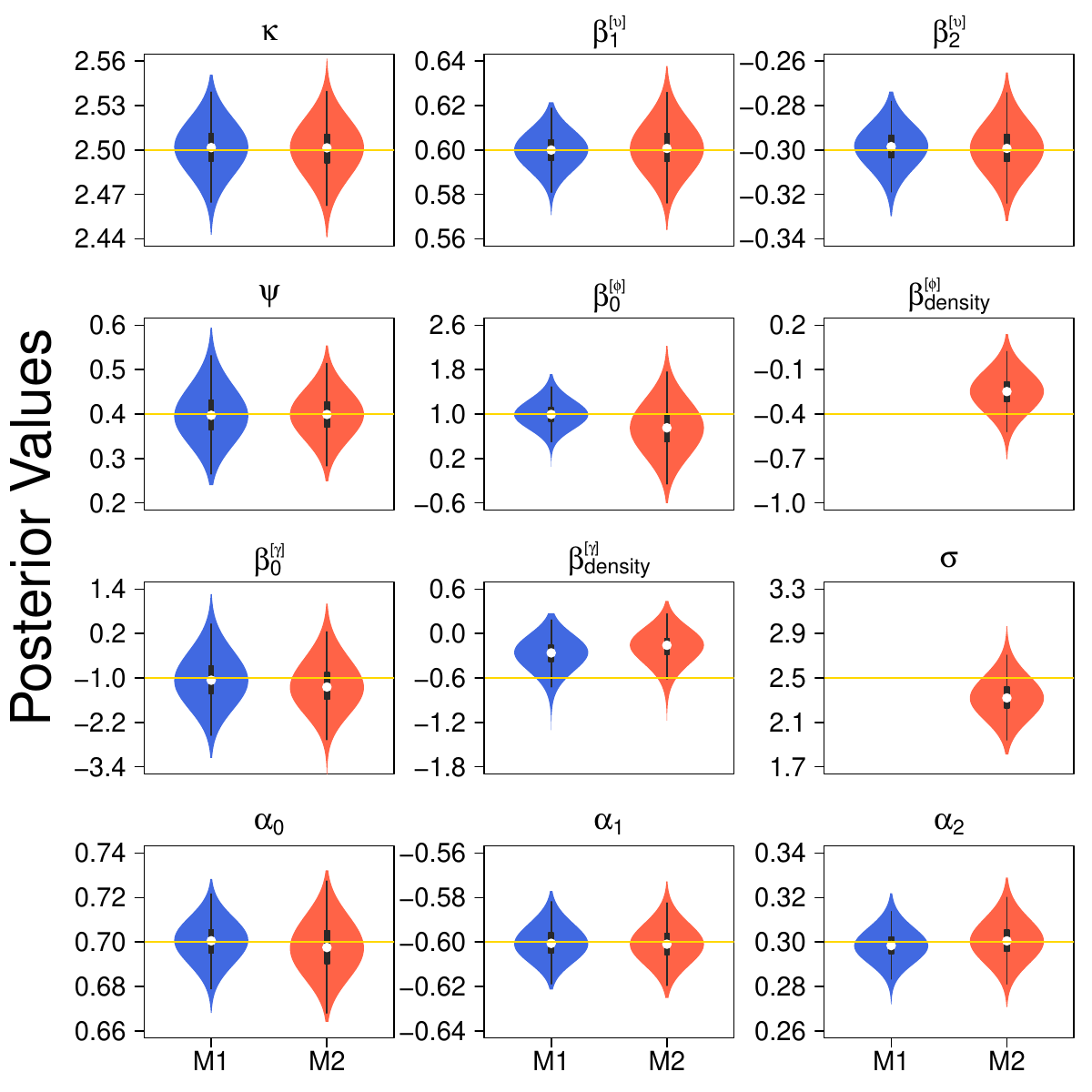}
\caption{\label{fig:sim}Violin plots showing the combined posterior samples from 20 simulations for the parameters of (M1) a SCR model that only allows density dependence on recruitment but not apparent survival and does not allow dispersal movement between years, and (M2) a SCR model that allows density dependence on both recruitment and apparent survival and dispersal movement.}
\end{figure}

The simulations showed that both the simpler model that only allows density dependence on recruitment but not apparent survival and does not allow dispersal movement between years and the more complex model that allows density dependence on both recruitment and apparent survival and dispersal movement between years can recover parameters related to habitat use ($\kappa$, $\beta^{[\upsilon]}$'s), initial existence ($\psi$) and detection ($\alpha$'s) without evident bias (Figure \ref{fig:sim}). However, both models underestimated the effects of density dependence on apparent survival ($\beta_{density}^{[\phi]}$) and recruitment ($\beta_{density}^{[\gamma]}$). The simpler model provided unbiased estimates for the mean of apparent survival ($\beta_0^{[\phi]}$) and recruitment ($\beta_0^{[\gamma]}$) at a logit scale, but the more complex model tended o have slightly biased estimates for these two parameters. The models also underestimated the standard deviation of Gaussian dispersal movement ($\sigma$).

\subsection{Real data}
\revisions{}{MCMC chains mixed well with Rhat statistics being 1.00 for all parameters of interest. Fig.~\ref{fig:tiger} showed posterior distributions of model parameters. Using our density dependent model, we found a \textit{positive} density dependence in survival $\phi$ at 90\% credible level. This might be a consequence of not having environmental covariates in the model.} 

\revision{Using our SCR model, we estimated parameters related to habitat use, density dependence apparent survival and recruitment, dispersal movement, and observation processes (Table.\ref{tab:tigerparameter}). In particular, we found little to no effect of density dependence on apparent survival ($\beta^{[\phi]}_{density}$: median -0.01, 80\% Credible Interval -0.13, 0.13), and a negative effect of density dependence on recruitment ($\beta^{[\gamma]}_{density}$: median -0.15, 80\% Credible Interval -0.32, -0.01).}

\begin{table}[]
    \centering
    \begin{tabular}{lcc}
    \hline
         Parameter& Median & 80\% CI  \\
    \hline
    $\kappa$    & 0.61& [0.53, 0.68]\\ 
    $\psi$    & 0.15& [0.09, 0.22]\\ 
    $\beta^{[\phi]}_0$    & 0.44& [-0.11, 1.08]\\ 
    $\beta^{[\phi]}_{\text{density}}$    & -0.01& [-0.13, 0.13]\\
    $\beta^{[\gamma]}_0$ &-1.04& [-2.02, -0.12]\\
    $\beta^{[\gamma]}_{\text{density}}$ &-0.15& [-0.32, -0.01]\\
    $\sigma$    & 2.89& [2.49, 3.35]\\
    $p$    & 0.51& [0.44, 0.59]\\
    \hline
    \end{tabular}
    \caption{Posterior medians and 80\% Credible Intervals of the parameters of the SCR model used in the tiger case study.\textbf{}}
    \label{tab:tigerparameter}
\end{table}

\section{Discussion}
In this study, we extended SCR models \citep[e.g.,][]{royle2013integrating,royle2018unifying} to estimate the true amount of time of habitat use, an information that is often of interest in ecological research and conservation planning \citep{hull2016habitat,dawson2013habitat}. \revisions{Using both simulations and a case study, we showed that it is possible to separate true habitat use from the observed number of captures under a hierarchical Bayesian Framework. A key component of our model is to allow habitat use be a function of both habitat heterogeneity and distance from territory centroid. Therefore, we highly recommend researchers to obtain and incorporate covariate information to maximize the utility of our model.}{In our simulations, we used the robust sampling design to facilitate the inference of habitat use. However, the inference of habitat use does not rely on robust sampling design, which is illustrated by our case study.} Another key component of our model is to assume that the total amount of time covered by the survey, $U$, is known, under the assumption of closed population within each primary period. \revisions{We recommend researchers to carefully select the primary period of their survey to meet this assumption. Further studies could also consider extending the model to allow the violation of this assumption due to, for example, death of an individual during the primary period.}{However, there could be cases when $U$ is unknown if the assumption of closed population is violated due to, for example, death of an individual during the primary period. We believe that the model performance should not be influenced if such a situation is rare or the variation in $U$ across individuals is small. It is also worthy to further explore the potential of estimating $U$ in the model, potentially with informative priors.}

\begin{figure}[htp]
\centering
\includegraphics[width=1.0\linewidth]{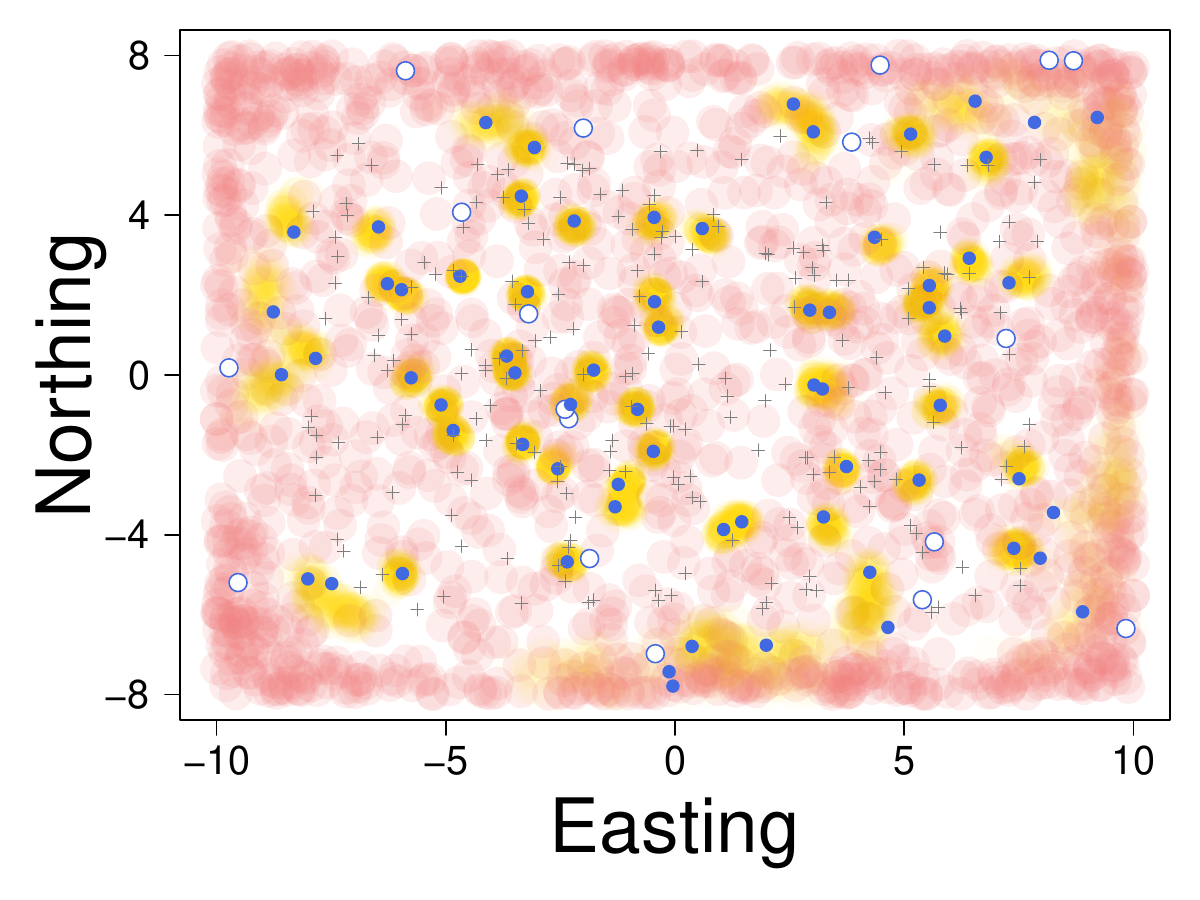}
\caption{\label{fig:ss}The true (blue solid point: observed individual, blue broken point: unobserved individual) and estimated territory centroids (yellow cloud: observed individual, red cloud: unobserved individual) in one year of a spatial capture-recapture study. Grey crosses represent the locations of camera traps.}
\end{figure}

Despite the accurate information of habitat use provided by this SCR model, we found it still difficult to estimate the effect of density dependence on apparent survival and recruitment with this model. So far, density dependence is mostly considered at the population level in SCR models \citep[e.g.,][]{chandler2014spatially}; when density dependence is considered at individual level in SCR models, only the locations of territory centroids, but not the information of habitat use, were considered \citep{milleret2023estimating}. The shortcoming of doing so is that individuals will not influence each other through density dependent processes if their territory centroids do not fall at the same site, despite the fact that their territories can still overlap with each other to induce competition. Therefore, we attempted to develop a model that links habitat use information with density dependence. The difficulties in estimating the effect of density dependence in this model is likely due to the fact that the current SCR model can only identify the locations of territory centroids of observed individuals, but not unobserved individuals without linking the locations of territory centroids to environmental covariates (Figure \ref{fig:ss}). Because the locations of unobserved individuals are misidentified, the model mis-specifies local density, and thus underestimates the effect of density dependence on vital rates. It is likely for the same reason that the model also provides biased estimates of the movement parameter. \revision{In the case study, our model identifies no density dependence on apparent survival and a negative density dependence on recruitment. Because of the model's tendency to underestimate density dependence, we cannot conclude that there is no density dependence on apparent survival, and studies have suggested that density dependent apparent survival is possible for tigers \citep{bisht2019demography}. On the other hand, the negative density dependence on recruitment seems to actually exit (although its strength may be underestimated), which indicates that the tigers may have a lower reproduction and/or immigration rate in areas used by more individuals. For instance, tigers are reported to be reluctant to move into a crowded area to prevent inbreeding depression \citep{smith1991contribution}.}

Overall, our results are in line with the findings of previous studies that showed the limits of the current generation of SCR models \citep{sun2024importance}, and it remains a challenge to link habitat use with vital rates at individual level in SCR models. \revision{Therefore, we suggest researcher taking cautions when using this and other SCR models to evaluate density dependence.} Further development of SCR models is needed to better identify the locations of individuals, probably through linking them to environmental covariates \citep[e.g.,][]{borchers2008spatially} or the integration of other data resources with SCR data \citep[e.g.,][]{sun2019incorporating}.

Our study nonetheless showed the importance and possibility of separating true habitat use with observation processes in SCR models. Our modeling approach can be used in a broad range of species and populations as long as SCR surveys are applicable. Our model can also be easily adapted to a variety of sampling designs \citep{dupont2021optimal,sun2014trap} or modified to include additional model components such as sex structures \citep{augustine2020sex}. Thus, our model is broadly transferable under the framework of spatial statistic trinity.\revisions{}{ In our case study, we found an unexpected positive density dependence, likely due to missing environmental covariates because higher density is likely to be an indication of better habitat and in turn higher survival rate. Indeed, including environmental covariates to inform habitat use is a key component of this model. Therefore, we highly recommend future studies to consider spatial location heterogeneity in habitat conditions in SCR models.}


\bibliographystyle{apalike}
\bibliography{references}

@article{dennis1994density,
  title={Density dependence in time series observations of natural populations: estimation and testing},
  author={Dennis, Brian and Taper, Mark L},
  journal={Ecological monographs},
  volume={64},
  number={2},
  pages={205--224},
  year={1994},
  publisher={Wiley Online Library}
}

@article{brook2006strength,
  title={Strength of evidence for density dependence in abundance time series of 1198 species},
  author={Brook, Barry W and Bradshaw, Corey JA},
  journal={Ecology},
  volume={87},
  number={6},
  pages={1445--1451},
  year={2006},
  publisher={Wiley Online Library}
}

@article{st1970detection,
  title={The detection of regulation in animal populations},
  author={St. Amant, JLS},
  journal={Ecology},
  volume={51},
  number={5},
  pages={823--828},
  year={1970},
  publisher={Wiley Online Library}
}

@article{hanski1990density,
  title={Density dependence, regulation and variability in animal populations},
  author={Hanski, Ilkka Aulis},
  journal={Philosophical Transactions of the Royal Society of London. Series B: Biological Sciences},
  volume={330},
  number={1257},
  pages={141--150},
  year={1990},
  publisher={The Royal Society London}
}

@article{turchin1992complex,
  title={Complex dynamics in ecological time series},
  author={Turchin, Peter and Taylor, Andrew D},
  journal={Ecology},
  volume={73},
  number={1},
  pages={289--305},
  year={1992},
  publisher={Wiley Online Library}
}

@article{royle2008hierarchical,
  title={A hierarchical model for spatial capture--recapture data},
  author={Royle, J Andrew and Young, Kevin V},
  journal={Ecology},
  volume={89},
  number={8},
  pages={2281--2289},
  year={2008},
  publisher={Wiley Online Library}
}

@article{efford2004density,
  title={Density estimation in live-trapping studies},
  author={Efford, Murray},
  journal={Oikos},
  volume={106},
  number={3},
  pages={598--610},
  year={2004},
  publisher={Wiley Online Library}
}

@article{borchers2008spatially,
  title={Spatially explicit maximum likelihood methods for capture--recapture studies},
  author={Borchers, David L and Efford, Murray G},
  journal={Biometrics},
  volume={64},
  number={2},
  pages={377--385},
  year={2008},
  publisher={Oxford University Press}
}

@article{royle2018unifying,
  title={Unifying population and landscape ecology with spatial capture--recapture},
  author={Royle, J Andrew and Fuller, Angela K and Sutherland, Christopher},
  journal={Ecography},
  volume={41},
  number={3},
  pages={444--456},
  year={2018},
  publisher={Wiley Online Library}
}

@article{ergon2014separating,
  title={Separating mortality and emigration: modelling space use, dispersal and survival with robust-design spatial capture--recapture data},
  author={Ergon, Torbj{\o}rn and Gardner, Beth},
  journal={Methods in Ecology and Evolution},
  volume={5},
  number={12},
  pages={1327--1336},
  year={2014},
  publisher={Wiley Online Library}
}

@article{gardner2010spatially,
  title={Spatially explicit inference for open populations: estimating demographic parameters from camera-trap studies},
  author={Gardner, Beth and Reppucci, Juan and Lucherini, Mauro and Royle, J Andrew},
  journal={Ecology},
  volume={91},
  number={11},
  pages={3376--3383},
  year={2010},
  publisher={Wiley Online Library}
}

@article{chandler2014spatially,
  title={Spatially explicit integrated population models},
  author={Chandler, Richard B and Clark, Joseph D},
  journal={Methods in Ecology and Evolution},
  volume={5},
  number={12},
  pages={1351--1360},
  year={2014},
  publisher={Wiley Online Library}
}

@article{milleret2023estimating,
  title={Estimating spatially variable and density-dependent survival using open-population spatial capture--recapture models},
  author={Milleret, Cyril and Dey, Soumen and Dupont, Pierre and Br{\o}seth, Henrik and Turek, Daniel and de Valpine, Perry and Bischof, Richard},
  journal={Ecology},
  volume={104},
  number={2},
  pages={e3934},
  year={2023},
  publisher={Wiley Online Library}
}

@article{royle2007analysis,
  title={Analysis of multinomial models with unknown index using data augmentation},
  author={Royle, J Andrew and Dorazio, Robert M and Link, William A},
  journal={Journal of Computational and Graphical Statistics},
  volume={16},
  number={1},
  pages={67--85},
  year={2007},
  publisher={Taylor \& Francis}
}

@article{lewis2023climate,
  title={Climate-mediated population dynamics of a migratory songbird differ between the trailing edge and range core},
  author={Lewis, William B and Cooper, Robert J and Chandler, Richard B and Chitwood, Ryan W and Cline, Mason H and Hallworth, Michael T and Hatt, Joanna L and Hepinstall-Cymerman, Jeff and Kaiser, Sara A and Rodenhouse, Nicholas L and others},
  journal={Ecological Monographs},
  volume={93},
  number={1},
  pages={e1559},
  year={2023},
  publisher={Wiley Online Library}
}

@misc{r2024r,
  title={R: A language and environment for statistical computing},
  author={R Core Team, R},
  year={2024},
  howpublish={Vienna, Austria}
}

@article{royle2013integrating,
  title={Integrating resource selection information with spatial capture--recapture},
  author={Royle, J Andrew and Chandler, Richard B and Sun, Catherine C and Fuller, Angela K},
  journal={Methods in Ecology and Evolution},
  volume={4},
  number={6},
  pages={520--530},
  year={2013},
  publisher={Wiley Online Library}
}

@article{pollock1982capture,
  title={A capture-recapture design robust to unequal probability of capture},
  author={Pollock, Kenneth H},
  journal={The Journal of Wildlife Management},
  volume={46},
  number={3},
  pages={752--757},
  year={1982},
  publisher={JSTOR}
}

@article{hassell1975density,
  title={Density-dependence in single-species populations},
  author={Hassell, Michael P},
  journal={The Journal of animal ecology},
  pages={283--295},
  year={1975},
  publisher={JSTOR}
}

@article{bellows1981descriptive,
  title={The descriptive properties of some models for density dependence},
  author={Bellows, TS},
  journal={The Journal of Animal Ecology},
  pages={139--156},
  year={1981},
  publisher={JSTOR}
}

@article{dennis2006estimating,
  title={Estimating density dependence, process noise, and observation error},
  author={Dennis, Brian and Ponciano, Jos{\'e} Miguel and Lele, Subhash R and Taper, Mark L and Staples, David F},
  journal={Ecological Monographs},
  volume={76},
  number={3},
  pages={323--341},
  year={2006},
  publisher={Wiley Online Library}
}

@article{herrando2012density,
  title={Density dependence: an ecological Tower of Babel},
  author={Herrando-P{\'e}rez, Salvador and Delean, Steven and Brook, Barry W and Bradshaw, Corey JA},
  journal={Oecologia},
  volume={170},
  pages={585--603},
  year={2012},
  publisher={Springer}
}

@article{ratikainen2008density,
  title={When density dependence is not instantaneous: theoretical developments and management implications},
  author={Ratikainen, Irja I and Gill, Jennifer A and Gunnarsson, T{\'o}mas G and Sutherland, William J and Kokko, Hanna},
  journal={Ecology Letters},
  volume={11},
  number={2},
  pages={184--198},
  year={2008},
  publisher={Wiley Online Library}
}

@article{travis2013evolution,
  title={Evolution in population parameters: density-dependent selection or density-dependent fitness?},
  author={Travis, Joseph and Leips, Jeff and Rodd, F Helen},
  journal={The American Naturalist},
  volume={181},
  number={S1},
  pages={S9--S20},
  year={2013},
  publisher={University of Chicago Press Chicago, IL}
}

@article{holt1996adaptive,
  title={Adaptive evolution in source-sink environments: direct and indirect effects of density-dependence on niche evolution},
  author={Holt, Robert D},
  journal={Oikos},
  pages={182--192},
  year={1996},
  publisher={JSTOR}
}

@article{howell2020informing,
  title={Informing amphibian conservation efforts with abundance-based metapopulation models},
  author={Howell, Paige E and Hossack, Blake R and Muths, Erin and Sigafus, Brent H and Chandler, Richard B},
  journal={Herpetologica},
  volume={76},
  number={2},
  pages={240--250},
  year={2020},
  publisher={Herpetologists' League}
}

@article{zhao2019integrated,
  title={Integrated modeling predicts shifts in waterbird population dynamics under climate change},
  author={Zhao, Qing and Boomer, G Scott and Royle, J Andrew},
  journal={Ecography},
  volume={42},
  number={9},
  pages={1470--1481},
  year={2019},
  publisher={Wiley Online Library}
}

@article{zabel2006interplay,
  title={The interplay between climate variability and density dependence in the population viability of Chinook salmon},
  author={Zabel, Richard W and Scheuerell, Mark D and McClure, Michelle M and Williams, John G},
  journal={Conservation Biology},
  volume={20},
  number={1},
  pages={190--200},
  year={2006},
  publisher={Wiley Online Library}
}

@article{sabo2004efficacy,
  title={Efficacy of simple viability models in ecological risk assessment: does density dependence matter?},
  author={Sabo, John L and Holmes, Elizabeth E and Kareiva, Peter},
  journal={Ecology},
  volume={85},
  number={2},
  pages={328--341},
  year={2004},
  publisher={Wiley Online Library}
}

@article{lamb2017influence,
  title={Influence of density-dependent competition on foraging and migratory behavior of a subtropical colonial seabird},
  author={Lamb, Juliet S and Satg{\'e}, Yvan G and Jodice, Patrick GR},
  journal={Ecology and Evolution},
  volume={7},
  number={16},
  pages={6469--6481},
  year={2017},
  publisher={Wiley Online Library}
}

@article{hixon2005competition,
  title={Competition, predation, and density-dependent mortality in demersal marine fishes},
  author={Hixon, Mark A and Jones, Geoffrey P},
  journal={Ecology},
  volume={86},
  number={11},
  pages={2847--2859},
  year={2005},
  publisher={Wiley Online Library}
}

@article{fronhofer2023shape,
  title={The shape of density dependence and the relationship between population growth, intraspecific competition and equilibrium population density},
  author={Fronhofer, Emanuel A and Govaert, Lynn and O'Connor, Mary I and Schreiber, Sebastian J and Altermatt, Florian},
  journal={Oikos},
  pages={e09824},
  year={2023},
  publisher={Wiley Online Library}
}

@article{detto2019bias,
  title={Bias in the detection of negative density dependence in plant communities},
  author={Detto, Matteo and Visser, Marco D and Wright, S Joseph and Pacala, Stephen W},
  journal={Ecology Letters},
  volume={22},
  number={11},
  pages={1923--1939},
  year={2019},
  publisher={Wiley Online Library}
}

@article{shenk1998sampling,
  title={Sampling-variance effects on detecting density dependence from temporal trends in natural populations},
  author={Shenk, Tanya M and White, Gary C and Burnham, Kenneth P},
  journal={Ecological monographs},
  volume={68},
  number={3},
  pages={445--463},
  year={1998},
  publisher={Wiley Online Library}
}

@article{judson1994rise,
  title={The rise of the individual-based model in ecology},
  author={Judson, Olivia P},
  journal={Trends in ecology \& evolution},
  volume={9},
  number={1},
  pages={9--14},
  year={1994},
  publisher={Elsevier}
}

@incollection{grimm2013individual,
  title={Individual-based modeling and ecology},
  author={Grimm, Volker and Railsback, Steven F},
  booktitle={Individual-based modeling and ecology},
  year={2013},
  publisher={Princeton university press}
}

@article{hooten2020statistical,
  title={Statistical implementations of agent-based demographic models},
  author={Hooten, Mevin and Wikle, Christopher and Schwob, Michael},
  journal={International Statistical Review},
  volume={88},
  number={2},
  pages={441--461},
  year={2020},
  publisher={Wiley Online Library}
}

@article{schafer2022bayesian,
  title={Bayesian inverse reinforcement learning for collective animal movement},
  author={Schafer, Toryn LJ and Wikle, Christopher K and Hooten, Mevin B},
  journal={The Annals of Applied Statistics},
  volume={16},
  number={2},
  pages={999--1013},
  year={2022},
  publisher={Institute of Mathematical Statistics}
}

@article{hull2016habitat,
  title={Habitat use and selection by giant pandas},
  author={Hull, Vanessa and Zhang, Jindong and Huang, Jinyan and Zhou, Shiqiang and Vina, Andres and Shortridge, Ashton and Li, Rengui and Liu, Dian and Xu, Weihua and Ouyang, Zhiyun and others},
  journal={PloS one},
  volume={11},
  number={9},
  pages={e0162266},
  year={2016},
  publisher={Public Library of Science San Francisco, CA USA}
}

@article{dawson2013habitat,
  title={Habitat use and conservation of an endangered dolphin},
  author={Dawson, Stephen and Fletcher, David and Slooten, Elisabeth},
  journal={Endangered Species Research},
  volume={21},
  number={1},
  pages={45--54},
  year={2013}
}

@misc{tigerdata,
    title={Ten year camera trap dataset of tigers in India [Dataset]},
    author = {Gardner, Beth and Sollmann, Rahel and Kumar, N. Samba and Jathanna, Devcharan and Karanth, K. Ullas},
    howpublished = {Dryad. \url{https://doi.org/10.5061/dryad.bcc2fqzd2}},
    note ={Accessed: 2024-11-20},
    year = {2020}
}

@article{karanth2006assessing,
  title={Assessing tiger population dynamics using photographic capture--recapture sampling},
  author={Karanth, K Ullas and Nichols, James D and Kumar, N Samba and Hines, James E},
  journal={Ecology},
  volume={87},
  number={11},
  pages={2925--2937},
  year={2006},
  publisher={Wiley Online Library}
}

@article{sun2024importance,
  title={The importance of independence in unmarked spatial capture--recapture analysis},
  author={Sun, Catherine and Cole Burton, A},
  journal={Wildlife Biology},
  volume={2024},
  number={3},
  pages={e01182},
  year={2024},
  publisher={Wiley Online Library}
}

@article{sun2019incorporating,
  title={Incorporating citizen science data in spatially explicit integrated population models},
  author={Sun, Catherine C and Royle, J Andrew and Fuller, Angela K},
  journal={Ecology},
  volume={100},
  number={9},
  pages={e02777},
  year={2019},
  publisher={Wiley Online Library}
}

@article{dupont2021optimal,
  title={Optimal sampling design for spatial capture--recapture},
  author={Dupont, Gates and Royle, J Andrew and Nawaz, Muhammad Ali and Sutherland, Chris},
  journal={Ecology},
  volume={102},
  number={3},
  pages={e03262},
  year={2021},
  publisher={Wiley Online Library}
}

@article{sun2014trap,
  title={Trap configuration and spacing influences parameter estimates in spatial capture-recapture models},
  author={Sun, Catherine C and Fuller, Angela K and Royle, J Andrew},
  journal={PloS one},
  volume={9},
  number={2},
  pages={e88025},
  year={2014},
  publisher={Public Library of Science San Francisco, USA}
}

@article{augustine2020sex,
  title={Sex-specific population dynamics and demography of capercaillie (Tetrao urogallus L.) in a patchy environment},
  author={Augustine, Ben C and K{\'e}ry, Marc and Olano Marin, Juanita and Mollet, Pierre and Pasinelli, Gilberto and Sutherland, Chris},
  journal={Population Ecology},
  volume={62},
  number={1},
  pages={80--90},
  year={2020},
  publisher={Wiley Online Library}
}

@book{zhao2024bayesian,
  title={Bayesian Analysis of Spatially Structured Population Dynamics},
  author={Zhao, Qing},
  year={2024},
  publisher={Springer Nature}
}

@article{broms2016dynamic,
  title={Dynamic occupancy models for explicit colonization processes},
  author={Broms, Kristin M and Hooten, Mevin B and Johnson, Devin S and Altwegg, Res and Conquest, Loveday L},
  journal={Ecology},
  volume={97},
  number={1},
  pages={194--204},
  year={2016},
  publisher={Wiley Online Library}
}

@article{zhao2020sampling,
  title={On the sampling design of spatially explicit integrated population models},
  author={Zhao, Qing},
  journal={Methods in Ecology and Evolution},
  volume={11},
  number={10},
  pages={1207--1220},
  year={2020},
  publisher={Wiley Online Library}
}

@article{efford2020spatial,
  title={A spatial open-population capture-recapture model},
  author={Efford, Murray G and Schofield, Matthew R},
  journal={Biometrics},
  volume={76},
  number={2},
  pages={392--402},
  year={2020},
  publisher={Oxford University Press}
}

@book{royle2013spatial,
  title={Spatial capture-recapture},
  author={Royle, J Andrew and Chandler, Richard B and Sollmann, Rahel and Gardner, Beth},
  year={2013},
  publisher={Academic press}
}

@book{efford2025secr,
  title={The SECR book: A handbook of spatially explicit capture–recapture
methods},
  author={Efford, Murray},
  year={2025},
  publisher={Self Publishing}
}

@article{bisht2019demography,
  title={Demography of a high-density tiger population and its implications for tiger recovery},
  author={Bisht, Shikha and Banerjee, Sudip and Qureshi, Qamar and Jhala, Yadavendradev},
  journal={Journal of Applied Ecology},
  volume={56},
  number={7},
  pages={1725--1740},
  year={2019},
  publisher={Wiley Online Library}
}

@article{smith1991contribution,
  title={The contribution of variance in lifetime reproduction to effective population size in tigers},
  author={Smith, James L David and McDougal, Charles},
  journal={Conservation Biology},
  volume={5},
  number={4},
  pages={484--490},
  year={1991},
  publisher={Wiley Online Library}
}

\end{document}